\documentclass[12pt,thmsa]{article}

\usepackage{amsfonts}
\usepackage{graphicx}
\usepackage{epsfig}
\usepackage{graphics}

\makeatletter
\newcommand{\row}[1]%
{\mathord{\buildrel{\lower3pt%
\hbox{$\scriptscriptstyle\rightarrow$}}\over #1}}

\newcommand{\dyadic}[1]{\mathord{\dyadic@rrow{#1}}}
\newcommand{\dyadic@rrow}[1]{
\begin{picture}(12,12)(-1,0)
\put(-3,12){\makebox(0,0)[t]{$\scriptscriptstyle\downarrow$}}
\put(-3,13){\makebox(0,0)[l]{$\scriptscriptstyle\longrightarrow$}}
\put(5,0){\makebox(0,0)[b]{$#1$}}
\end{picture}
}

\newcommand{\bra}[1]{\bigl\langle #1 \bigr|}
\newcommand{\ket}[1]{\bigl| #1 \bigr\rangle}

\begin{document}
\begin{center}
{\large Entangled Network and Quantum Communication }

N. Metwally \\[0pt]

Math. Dept., Faculty of Science, South Valley University, Aswan, Egypt \\[0pt]
E.mail: Nmetwally$@$gmail.com
\end{center}
 \begin{abstract}
A theoretical scheme is introduced to generate entangled network
via  Dzyaloshinskii- Moriya (DM)interaction. The dynamics of
entanglement  generated between  different nodes by direct or
indirect interaction is investigated. It is shown that, the
direction of (DM) interaction  and the location of the nodes have
a sensational effect on the degree of entanglement. We quantify
the minimum  entanglement  generated between all the nodes. The
upper and lower bound of the entanglement of the generated network
depends on the direction of  DM interaction and the repetition of
the behavior depends on the strength of DM. The generated
entangled nodes are used as quantum channel to perform quantum
teleportation, where we show that the fidelity of teleporting
unknown information between the network members depends on the
location of the members.

\bigskip

\end{abstract}

\section{Introduction}
Quantum entanglement is considered as a promising resource for
quantum information and computation fields\cite{nil}. Long
surviving  generated entangled states  is an important issue in
the context of quantum information processing\cite{khs}.
Generating entangled state between different types of bipartite
systems have been extensively investigated \cite{Zhen}.
Multi-parties entanglement is more powerful  than bipartite
entanglement in quantum information processing.  For example, M.
Siomau and et. al. \cite{Soi} discussed the entanglement evolution
of  multi-qubit systems when one of its qubits is subjected to a
general noisy channel.  Multipartite entangled states with two
bosonic modes and qubits have been discussed by Munhoz and
Semi$\tilde{a}$o \cite{Mun}. Perseguers and et.al. \cite{Per} have
studied  the problem of creating a long-distance entangled state
between two stations of a network. A general scheme for
construction of noiseless networks detecting entanglement with the
help of linear, hermiticity-preserving maps have been introduced
in \cite{Hor}.

In this contribution, we introduce a theoretical technique  to
generate entangled network by using pairs of maximum entangled
states. The description of this scheme is shown in Fig.1, where it
is assumed that a source supplies the users with pairs of Bell
states. The second and third qubits entangle together via
Zyaloshinskii- Moriya (DM) interaction \cite{Moriya}. This type of
interaction is very important in the context of quantum
information and entanglement, where Chutia  and et.al \cite{Chut}
have  proposed an experiment using coupled quantum dots to detect
and characterize the DM interaction. Also in \cite{Jout} the
Dzyaloshinskii–Moriya interaction has been detected by means of
pulsed EPR spectroscopy.

Due to this interaction, all the qubits which represent  nodes in
the network are entangled together(the details are given in
Sec.2). We investigated the dynamics of entanglement  generated
between the different nodes  interacting directly or indirectly.
The possibility of using these channels to perform quantum
teleportation. is discussed

The paper is organized as follows In Sec.2, we introduce the
system and its analytical solution. The dynamics of the
entanglement between different nodes is investigated in Sec.3.1.
The upper and lower bounds of entanglement for the generated
entangled network is quantified in Sec.3.2. We introduce an
extension to the network to include more nodes in Sec.4.
 Employing the entangled channel
between the different nodes to achieve quantum teleportion is
discussed in Sec.5. Finally we summarize our results in Sec.6.

\section{The System}

Let us assume  that we have a source suppling the users with  two
qubit  pairs   prepared  in one of the  Bell (EPR)states,
$\ket{\phi^{\pm}}=\frac{1}{\sqrt{2}}(\ket{11}\pm\ket{00}),
\ket{\psi^{\pm}}=\frac{1}{\sqrt{2}}(\ket{10}\pm\ket{01})$\cite{EPR}.
For a convenience, these EPR states are  described by  Pauli
matrices. For example, the density operator
$\rho_{\phi^+}=\ket{\phi^+}\bra{\phi^+}$ takes the following form

\begin{equation}\label{2Q}
\rho_{\phi^+}=\frac{1}{4}(1+\sigma_x\tau_x-\sigma_y\tau_y+\sigma_z\tau_z),
\end{equation}
where, the vectors $\row{\sigma_i}=(\sigma_x,\sigma_y,\sigma_z)$
and $\row{\tau_j}=\tau_x,\tau_y,\tau_z)$ are Pauli operators (see
for example\cite{Metwally}). Each qubit is sent to two distinct
partners, which represent nodes in the network. These nodes are
connected together via Dzyaloshinskii- Moriya (DM) interaction,
where the end of each entangle node interacts with the first node
of the other entangled two nodes. To clarify this suggested
network, we start with four nodes network. Therefore, the initial
state vector of the network is given by
\begin{equation}\label{in}
\rho_{1234}(0)=\rho_{\phi^{+}_{12}}\otimes\rho_{\phi^{+}_{34}},
\end{equation}
where $\rho_{\phi_{12}^+}$ and $\rho_{\phi_{34}^+}$,
\begin{eqnarray}\label{2Q}
\rho_{\phi_{12}^+}&=&\frac{1}{4}(1+\sigma^{(1)}_x\tau^{(2)}_x-\sigma^{(1)}_y\tau^{(2)}_y+\sigma^{(1)}_z\tau^{(2)}_z),
\nonumber\\
\rho_{\phi_{34}^+}&=&\frac{1}{4}(1+\sigma^{(3)}_x\tau^{(4)}_x-\sigma^{(3)}_y\tau^{(4)}_y+\sigma^{(3)}_z\tau^{(4)}_z).
\end{eqnarray}
The  nodes $2$ and $3$ are connected via  DM interaction, which is
defined by,
\begin{equation}\label{DM}
H_{DM}=\row{D}\cdot(\row{\sigma_i}\times\row{\tau_j}).
\end{equation}
The components of the vector  $\row{D}=(D_x,D_y,D_z)$ are the
strength of $DM$ interaction in the directions of $x,y$ and $z-$
axis \cite{Moh}. In this treatment, we consider that DM
interaction is switched on the $z-$ or $x-$ axis. The time
evolution of the initial network (\ref{in}) is given by
\begin{equation}
\rho_{1234}(t)=\mathcal{U}_z(t)\rho_{1234}(0){\mathcal{U}^\dagger_{z}(t)},
\end{equation}
where $\mathcal{U}_{z}(t)$  represents the unitary operator when
DM interaction is switched  on the $z$-axis. In terms of Pauli
operators, it takes the form,
\begin{figure}[t!]
  \begin{center}
 \includegraphics[width=15pc,height=12pc]{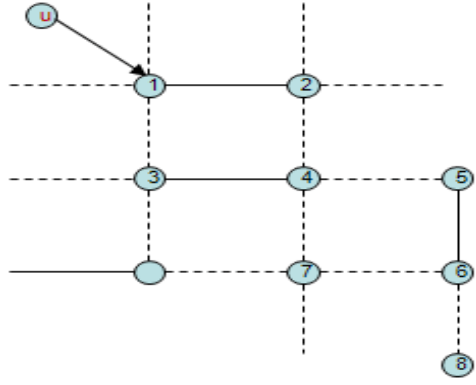}
 \end{center}
\caption{\small{The suggested network consists of a pair of
maximum entangled nodes of Bell type which is represented by the
solid line as $\rho_{12}$ and $\rho_{34}$. The dot lines represent
the generated entangled nodes via direct or indirect interaction.
 The qubit $u$, is unknown information is
given to the node $1$ who wish to teleport to any another node on
the network.}}
\end{figure}

\begin{equation}
\mathcal{U}^{(23)}_{z}(t)=cos^2(D_z t)+\sin^2(D_z
t)\tau^{(2)}_{z}\sigma_{z}^{(3)}-\frac{i}{2}\sin (2D_z
t)(\tau^{(2)}_x\sigma^{(3)}_y-\tau^{(2)}_y\sigma^{(3)}_x),
\end{equation}
where we assume that the interaction is running between  qubits
$2$ and $3$. Since, we are interested in investigating the
properties of the  channels between the  nodes, we calculate  the
density operator for each subsystem in this network. For example,
the density operator between the nodes  $1$ and $2$ is given by
$\rho_{12z}=tr_{34}\{\rho_{1234z}(t)\}$,
\begin{equation}\label{12t}
\rho_{12z}(t)=\frac{1}{4}(1+c^{(12)}_{xx}\sigma_x^{(1)}\tau_x^{(2)}-c^{(12)}_{yy}\sigma_y^{(1)}\tau_y^{(2)}
+c^{(12)}_{zz}\sigma_z^{(1)}\tau_z^{(2)}),
\end{equation}
where $ c^{(12)}_{xx}=-c^{(12)}_{yy}=\cos^4(D_z t)-\sin^{4}(D_z
t)$ and $ c^{(12)}_{zz}= \bigl(\cos^4(D_z t)+\sin^{4}(D_z
t)\bigr)-\sin^2(2D_zt)$. It is clear that the state (\ref{12t}) is
no longer maximum entangled state and it is of Werner type
\cite{Werner}. This means that due to the interaction with its
neighbor entangled two nodes, the initial state
$\rho_{\phi^+_{12}}$ loose some of its entanglement. Also, the
density operator between the nodes $"1"$ and $"3"$ ($\rho_{13}$)
is defined by
\begin{equation}
\rho_{13z}=\frac{1}{4}(1-c^{(13)}_{xy}\sigma^{(1)}_x\tau^{(3)}_y+c^{(13)}_{yx}\sigma^{(1)}_y\tau_x^{(3)}+c^{(13)}_{zz}\sigma^{(1)}_z\tau^{(3)}_z),
\end{equation}
where $ c^{(13)}_{xy}=c^{(13)}_{yx}=\frac{i}{4}\cos^2([D_z
t)\sin(2 D_zt), c^{(13)}_{zz}=\frac{1}{2}(1+\frac{1}{2}\sin^2(2
D_z t))$. Similarly, the density operator between the nodes $"1"$
and $"4"$, $\rho_{14z}=tr_{23}\{\rho_{1234z}(t)\}$ is defined by
\begin{equation}
\rho_{14z}(t)=\frac{1}{4}(1+c^{(14)}_{xy}\sigma_x^{(1)}\tau^{(4)}_y-c^{(14)}_{yx}\sigma^{(1)}_y\tau^{(4)}_x+
c^{(14)}_{zz}\sigma^{(1)}_z\tau^{(4)}_z),
\end{equation}
where $ c^{(14)}_{xy}=c^{(14)}_{yx}=i\sin^2(2D_zt)\cos^2D_z t,~
c^{(14)}_{zz}=\frac{1}{2}\bigl[1-\cos(2D_zt)-4\sin^2(2D_z
t)\bigr]. $ The density operator between the systems $2$ and $3$,
which represent the direct interaction system is given by
\begin{equation}
\rho_{23z}(t)=\frac{1}{4}(1+c^{(23)}_{xx}\sigma^{(2)}_x\tau^{(3)}_x+c^{(23)}_{yy}\sigma^{(2)}_y\tau^{(3)}_y+
c^{(23)}_{xy}\sigma_x^{(2)}\tau^{(3)}_x-c^{(23)}_{yx}\sigma^{(2)}_y\tau^{(3)}_x+
c^{(23)}_{zz}\sigma^{(2)}_z\tau^{(3)}_z),
\end{equation}
where,
 \begin{eqnarray} c^{(23)}_{xx}&=&c^{(23)}_{yy}=\sin(2D_z
t)\Bigl[\frac{1}{2}\sin^2(D_z t)-\frac{3}{4}\cos^2(D_z t)\Bigr],
\nonumber\\
c^{(23)}_{xy}&=&c^{(23)}_{yx}=\frac{i}{2}\sin(2D_z
t)\Bigl[1-\frac{1}{2}\cos^2(D_z t)\Bigr],
\nonumber\\
c^{(23)}_{zz}&=&\frac{9}{2}\sin^2(2D_z t).
\end{eqnarray}
The generated entangled channel between the nodes $"2"$ and $"4"$
is defined by
\begin{equation}
\rho_{24z}=\frac{1}{4}(1+c^{(24)}_{zz}\sigma^{(2)}_z\tau^{(4)}_z),
\end{equation}
where $c^{(24)}_{zz}=\frac{1}{2}+\frac{1}{8}\sin(2 D_z
t)^2-\frac{1}{2}\cos^2(2D_z t)$.

 Let us consider that the DM interaction  is switched on
 the $x-$ axis. In this case, the unitary operator is described by,
 \begin{equation}
\mathcal{U}^{(23)}_x(t)=cos^2(D_x t)+\sin^2(D_x
t)\tau^{(2)}_{x}\sigma_{x}^{(3)}-\frac{i}{2}\sin (2D_x
t)(\tau^{(2)}_z\sigma^{(3)}_y-\tau^{(2)}_y\sigma^{(3)}_z),
\end{equation}
where $D_x$ represents the strength of the interaction in  the
$x-$ direction. It is possible to obtain the density operators for
all the quantum channels which are generated by the direct or
indirect interaction. For example, the entangled quantum channel
between the nodes $1$ and $2$, $\rho_{12x}(t)$ evolves as
\begin{equation}
\rho_{12x}=\frac{1}{4}(1+c^{(12)}_{xx}\sigma^{(1)}_x\tau^{(2)}_x+c^{(12)}_{yy}\sigma^{(2)}_y\tau^{(2)}_y
+c_{zz}^{(12)}\sigma_z^{(1)}\tau_z^{(2)}),
\end{equation}
where $c^{(12x)}_{xx}=1-\frac{3}{2}\sin^2 (2 D_x t),~
 c^{(12x)}_{yy}=-\bigl(\cos^4( D_x t)-\sin^4(D_x t)\bigr)$ and $c^{(12x)}_{zz}=-c^{(12x)}_{yy}$. Similarly one
 obtains the density operators for the other subsystems. As a
direct interaction, we consider the density operator between the
second and third node, $\rho_{23x}=tr_{14}\{\rho_{1234}\}$ which
takes the form,
\begin{equation}
\rho_{23x}=\frac{1}{4}(1+c^{(23x)}_{xx}\sigma^{(2)}_x\tau^{(3)}_x+c^{(23x)}_{yy}\sigma^{(2)}_y\tau^{(3)}_y),
\end{equation}
where $c^{(23x)}_{xx}=-\frac{1}{2}\sin^2(2D_x t),~
c^{(23x)}_{yy}=\frac{5}{4}\sin^2(2D_x t)$.

\section{ Network Correlation}
\subsection{ Bi-partite entanglement}

In preceding section, we showed that there are some new channels
that  have been generated due to the indirect interaction. The
initial entangled nodes  are no longer maximum entangled. In this
section, we investigate the
 entanglement dynamics of  the initial
entangled nodes and quantify how much of entanglement survive
between them. Also, we quantify the amount of entanglement which
is generated between the nodes via indirect interaction. The
simplest way to do this is to use Wootters concurrence as a
measure of the degree of entanglement \cite{wootter}. A two-qubit
entanglement is quantified by the concurrence, whose definition is
given by
\begin{equation}
\mathcal{C}=max\bigl\{\sqrt{\lambda_1}-\sqrt{\lambda_2}-\sqrt{\lambda_3}-{\sqrt\lambda_4},0\bigr\},
\end{equation}
where $\lambda_1\geq\lambda_2 \geq\lambda_3 \geq\lambda_4$ are the
square roots of the eigenvalues of $\bar\rho\rho$. The density
operator $\rho$ represents the reduced density operator of the
total system and $\bar\rho=\sigma_y\tau_y\rho^*\tau_y\sigma_y$
with $\rho^*$ is the complex conjugate of $\rho$.

\begin{figure}
  \begin{center}
 \includegraphics[width=25pc,height=15pc]{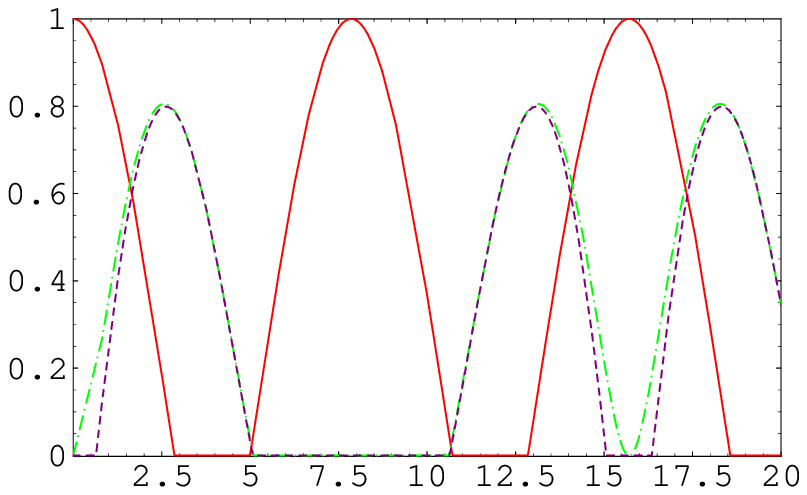}~
\put(-130,-10){\large $t$}
 \put(-318,102){\large $\mathcal{C}_{ijz}$}
  \end{center}
\caption{The dynamics of entanglement between different nodes. The
solid, dash dot and dot curves represent the dynamics of the
concurrence $\mathcal{C}_{ijz}, ij=12,13,14$ respectively. The  DM
interaction is switched on the $z-$axis with strength $D_z=0.2$.}
\end{figure}

 In Fig.(2), we plot the dynamics of the concurrence $\mathcal{C}_{ij}$ between the nodes $i$ and $j$ for the entangled
 channels $\rho_{ijz},ij=12,13,14$. The solid curve
represents the dynamics of the concurrence  $\mathcal{C}_{12z}$
for the channel $\varrho_{12z}(t)$. It is clear that at $t=0$, the
 entanglement is maximum i.e., $\mathcal{C}_{12z}=1$, since the
initial state of the nodes $"1"$ and $"2"$,
$\varrho_{12z}(0)=\rho_{\phi^{+}}$. However as $t$ increases,
$\mathcal{C}_{12}$ fluctuates between the maximum value  and a
minimum value, $\mathcal{C}_{12z}\simeq 0.25$). Due to the
interaction between the second and third nodes of the network
there are different entangled channels   generated between the
other nodes. For example, the  entanglement which is generated
between the node $"1"$ and $"3"$ is quantified by
$\mathcal{C}_{13z}$( dash-dot curve). Since at $t=0$, the nodes
$"1"$ and $"3"$ are completely separable, hence the degree of
entanglement ($\mathcal{C}_{13z})=0$. However, as the interaction
is switched on an entangled channel is generated between these
nodes. The degree of entanglement increases to reach its maximum
value $(\mathcal{C}_{13z}\simeq 0.79)$. For larger time, the
concurrence
 decreases to reach its minimum value
($\mathcal{C}_{13z}=0$). This means that the two nodes become
separable  while the state $\rho_{12z}$ turns into  a maximum
entangled state.

\begin{figure}
  \begin{center}
   \includegraphics[width=25pc,height=15pc]{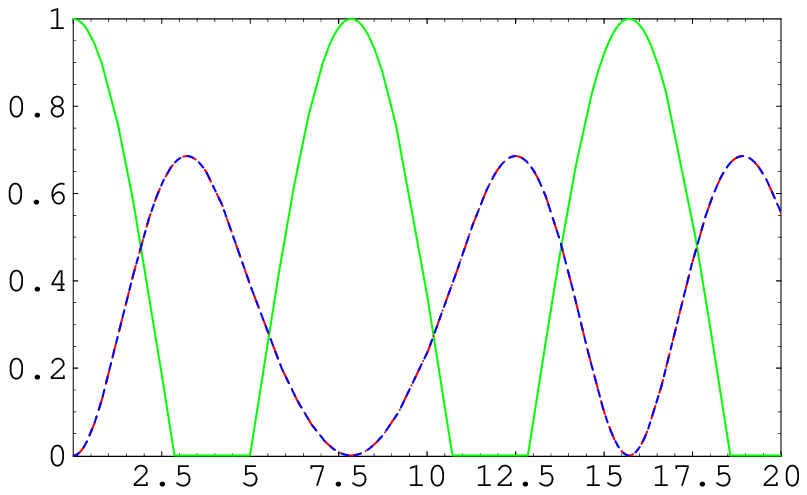}
\put(-130,-10){\large $t$}
 \put(-318,102){\large $\mathcal{C}_{ijz}$}
  \end{center}
\caption{The same as Fig.(1) but for the channels
$\rho_{ij},ij=34,23,24$ respectively.}
\end{figure}

Finally  the  behavior of $\mathcal{C}_{14z}$ which represents the
entanglement  between the nodes $"1"$ and $"4"$ shows that the
entangled channel  is not generated as soon as the interaction is
switched on.
 Also, the general behavior of $\mathcal{C}_{14z}$ is almost the same as that depicted for
$\mathcal{C}_{13z}$. This shows that with an equal probability one
can generate entangled channels between the node $"1"$ and ($"3"$
or $"4"$) with almost the same degree of entanglement. Therefore,
it is possible to send information from node $"1"$ to $"3"$ or
$"4"$ with the same fidelity.
\begin{figure}
  \begin{center}
 \includegraphics[width=25pc,height=15pc]{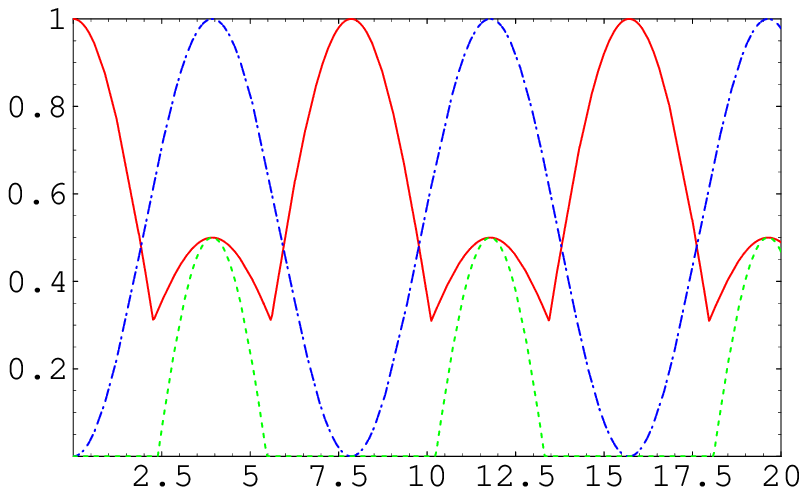}
 \put(-130,-10){\large $t$}
 \put(-318,102){\large $\mathcal{C}_{ijx}$}
\end{center}
\caption{The same as Fig.1 but DM is considered in $x-$ axis with
$D_x=0.2$. }
\end{figure}

Fig.(3) displays the dynamics of the concurrence
$\mathcal{C}_{ijz}$, $ij=23,24,34$, which measure the entanglement
in the channels $\rho_{23z},\rho_{24z},\rho_{34z}$ respectively .
The behavior of $\mathcal{C}_{34}$ (solid curve) which represents
the entanglement between the nodes $"3"$ and $"4"$, is the same as
that shown for the concurrence $\mathcal{C}_{12z}$(solid curve in
Fig.1). Also, the dynamics of entanglement between the nodes $"2"$
and $"3"$ is given by the concurrence $\mathcal{C}_{23z}$( dot
curve).
 For $t>0$, $\mathcal{C}_{23z}$ increases  smoothly  and
 reaches its upper bound  for the first time at $t\simeq 4$. For
  $t>4$, the entanglement decreases smoothly to
 vanish completely  for the first time at $t\simeq 7.5$. This behavior is repeated
 depending on the value of the interaction's strength.  At this
 time all the correlation between the other nodes is almost zero
 and consequently the network turns into its initial state.
 As an important observation, the entangled channel between the
 nodes $"2"$ and $4"$  generated via indirect interaction $\mathcal{C}_{24z}$
 is the same as that generated via direct interaction  between the nodes $"2"$ and
 $3"$, namely $\mathcal{C}_{23z}$=$\mathcal{C}_{24z}$ as shown in
 Fig.(3).

 Fig.(4), shows the effect of a different direction of  DM interaction, where we assume
 that it is switched on the  $x-$ axis. In this case, the dynamics
 of entanglement for the  initial entangled nodes  which is
 represented by $\mathcal{C}_{12x}$, $\mathcal{C}_{34x}$ for
 the channels $\rho_{12x}$ and  $\rho_{34x}$ respectively are
 completely different from those displayed in Figs.(2$\&3$). For
 example,  as soon as the interaction is
 switched on, $\mathcal{C}_{12x}$ decreases smoothly but it
 doesn't vanish. Also, the entangled channel between the nodes
 $"1"$ and $"3"$ turns into maximum entangled channel at $t\simeq 4$
 namely $\mathcal{C}_{13x}=1$, while  the maximum value of  $\mathcal{C}_{13z}\simeq
 0.79$.  The degree of entanglement between the nodes
 $"1"$ and $4"$ which is quantified by $\mathcal{C}_{14x}$ is
 different from the behavior of $\mathcal{C}_{14z}$, where it is
 generated after a longer time from the beginning of the
 interaction. The dynamics of entanglement $\mathcal{C}_{23x}$ which is generated between the nodes $2"$ and $"3"$
  via a  direct interaction is similar to that shown in Fig(3),
  but its maximum value is smaller than $\mathcal{C}_{23z}$. As a
  final observation, the dynamics of the entanglement which is
  generated between the nodes $"2"$ and $"4"$ is the same as that
  depicted between $"1"$ and $"3"$, namely
  $\mathcal{C}_{13x}=\mathcal{C}_{24x}$.

  From our finding, it is possible to generate entangled channel
  between the different nodes. The direction of the interaction
  plays an important role  in generating maximum entangled
  channel.

\begin{figure}
  \begin{center}
  \includegraphics[width=25pc,height=15pc]{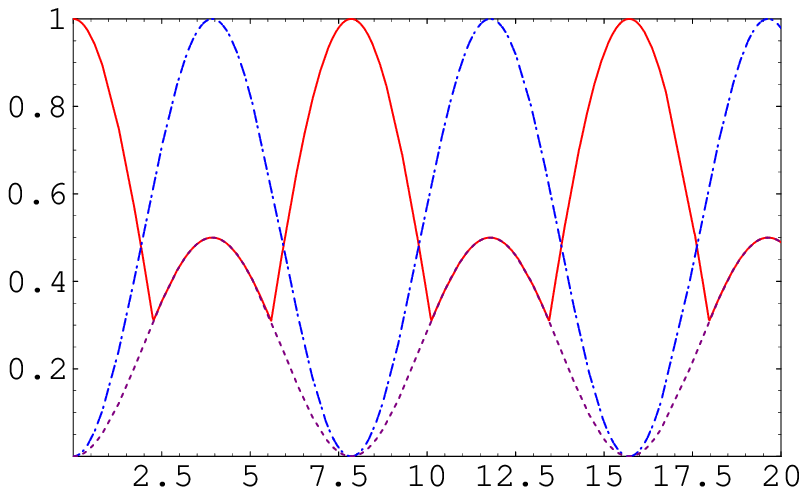}
\put(-130,-10){\large $t$}
 \put(-318,102){\large $\mathcal{C}_{ijx}$}
 \end{center}
\caption{The same as Fig.1 but DM is considered in $x-$ axis with
$D_x=0.2$. }
\end{figure}

\subsection{ Minimum entanglement of the Network}

In this subsection, we quantify the minimum amount of entanglement
contained in the generated entangled network. We use a measure
given by the concurrence introduced in \cite{Soi,Fei}. It  states
that, for a given pure N-qubit state $\ket{\psi}$, the concurrence
is defined by
\begin{equation}
\mathcal{C}_{min}=\sqrt{1-\frac{1}{N}\sum_{i=1}^{N}Tr\rho_i^2},
\end{equation}
where, $\rho_i=tr{\ket\psi\bra\psi}$ is the reduced  density
operator of the i-th qubit which is obtained by tracing out the
remaining $N-1$ qubits.
\begin{figure}
  \begin{center}
 \includegraphics[width=25pc,height=15pc]{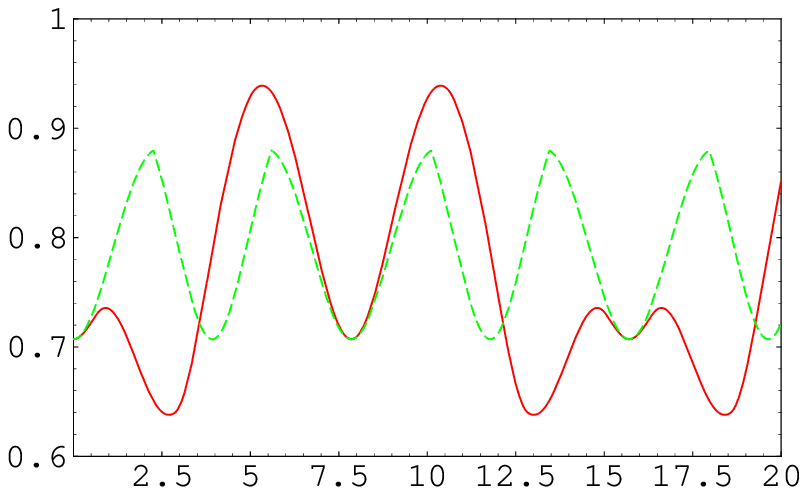}~
 \put(-140,-10){\large $t$}
 \put(-310,85){\large $\mathcal{C}_{min}$}
\end{center}
\caption{The minimum entanglement of the four qubit network,
$\mathcal{C}_{min}$ where  DM  interaction is switched in the $z-$
axis (solid curve) and in the $x$-axis (dot curve) $D_z=D_x=0.2$.
}
\end{figure}

Fig.(4), shows the dynamics of the minimum amount of entanglement
for the four qubit entangled network, where DM interaction  is
switched on the  $z-$ and $x-$ axis. It is clear that at $t=0$,
the concurrence $\mathcal{C}_{min}>0.7$ which represents  the
lower bound of entanglement. For $t>0$, the dynamics of
concurrence depends on the direction of the interaction. It is
clear  that when DM is switched on the  $z-$axis,
$\mathcal{C}_{min}$ reaches  its minimum value for the first time
at $t\simeq 2.6$. On the other hand, if DM is switched on the
$x-$axis, the dynamics of $\mathcal{C}_{min}$ is more stable,
where it oscillates  between a fixed  maximum and minimum values.

Therefore, it is possible to generate entangled network by using
pairs of EPR interact together via DM interaction. The upper and
lower bounds of the entanglement of the generated network depends
on the direction DM interaction and the repatation of the behavior
depends on the strength of DM interaction.

\section{Generalization of the Network}
In this section, we extend the entangled network (5) to include
more nodes. For this aim, we assume that according to the
preceding procedure which is described in Sec.2, there is an
entangled network consists of $4$ qubits  interact with another
two qubit state, $\rho_{56}$  defined as
\begin{equation}
\rho_{\phi_{56}^+}=\frac{1}{4}(1+\sigma^{(5)}_x\tau^{(6)}_x-\sigma^{(5)}_y\tau^{(6)}_y+\sigma^{(5)}_z\tau^{(6)}_z).
\end{equation}
 The time evaluation of the final state is given by,
\begin{equation}
\rho_{1...6}(t)=\mathcal{U}_{45z}(t)\rho(t)_{1..4}\otimes\rho_{\phi^{+}_{56}}\mathcal{U}^{\dagger}_{45z},
\end{equation}
where, $\mathcal{U}_{45z}$ represents the unitary operator when DM
interaction  is switched in the $z-$axis and the interaction is
running between the
qubits $"4"$ and $5"$. %

 To show the dynamics of entanglement between different nodes, we have to
obtain the reduced density operator of the required subsystem by
tracing out the other  subsystems. For example, the quantum
channel between the nodes $"1"$ and $"5"$ is represented by the
density operator
$\rho_{15}=tr_{2346}\Bigl\{\ket{\psi}_{1...6}\bra{\psi}\Bigr\}$.
In the computational basis one can rewrite this density operator
as
\begin{eqnarray}
\rho_{15}&=&\frac{1}{4}\Bigl\{1+(\mu_1+\mu_2-\mu_3-\mu_4)\sigma_z^{(1)}+(\mu_1+\mu_3-\mu_2-\mu_4)\tau_z^{(1)}+
+i\mu_5(\sigma^{(1)}_x\tau^{(5)}_y-\sigma^{(1)}_y\tau^{(5)}_x)
\nonumber\\
&& \hspace{6cm}+\mu_6\sigma_z^{(1)}\tau_x^{(5)}
+(\mu_1+\mu_4-\mu_2-\mu_3)\sigma^{(1)}_z\tau^{(5)}_z\Bigr\},
\end{eqnarray}
\begin{figure}
  \begin{center}
 \includegraphics[width=25pc,height=15pc]{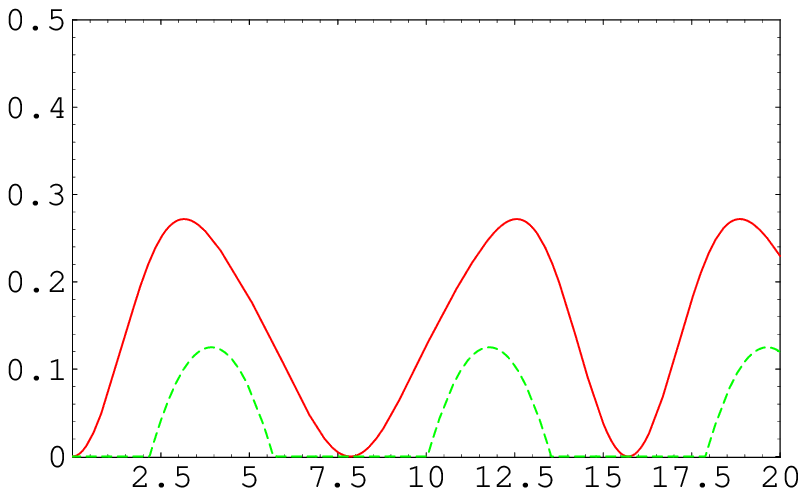}~
 \put(-140,-10){\large $t$}
 \put(-310,90){\large $\mathcal{C}_{ij}$}
\end{center}
\caption{The dynamics of the concurrence $\mathcal{C}_{ij},
ij=15,16$ between the nodes $"1"\&"5"$ (solid curve ) and
"$1"\&"6"$(dot curve) where DM interaction is switched in the
$z-$ axis,) $D_z=D_x=0.2$. }
\end{figure}
where
\begin{eqnarray}
\mu_1&=&\frac{3+\mathcal{S}^2}{8},\quad
\mu_2=\frac{1+\mathcal{S}^2}{4}, \quad
\mu_3=\frac{1}{8}\bigl[1+\mathcal{S}^2(\mathcal{C}^2+\mathcal{C})+\mathcal{C}^2\bigr],
\nonumber\\
\mu_5&=&\frac{1}{4}\mathcal{C}\mathcal{S}^2,\quad\quad
\mu_6=\frac{1}{4}\bigl[\mathcal{S}^2(1+\mathcal{C}^2)+\mathcal{C}^4\bigr],
\end{eqnarray}
In a similar way the density operator between the nodes $"1"$ and
$"6"$ is given by

\begin{eqnarray}
\rho_{16}&=&\frac{1}{4}\Bigl\{1+(\nu_1+\nu_4-\nu_2-\nu_3)\sigma_z^{(1)}+(\nu_1+\nu_3-\nu_2-\nu_4)\tau_z^{(1)}+
\nu_5(\sigma^{(1)}_x\tau^{(5)}_x-\sigma^{(1)}_y\tau^{(5)}_y)
\nonumber\\
 &&\hspace{4cm} +
(\nu_1+\nu_2-\nu_3-\nu_4)\sigma^{(1)}_z\tau^{(5)}_z\Bigr\},
\end{eqnarray}
where
\begin{eqnarray}
\nu_1&=&\frac{1+\mathcal{S}^2}{4},\quad \nu_2=\frac{1}{8},\quad
\nu_3=\frac{1}{8}\bigl[2\mathcal{C}^2+\mathcal{S}^2(1+\mathcal{C}^2\bigr],
\nonumber\\
\nu_4&=&\frac{1}{8}\bigl[4+\mathcal{S}^2+\mathcal{S}^4\bigr],\quad
\nu_5=\frac{1}{8}\mathcal{C}^2\mathcal{S}^2,
\end{eqnarray}
where $\mathcal{S}=\sin(2D_z t), \mathcal{C}=\cos(2D_z t)$.

Fig.7, shows the dynamics of the concurrence for the quantum
entangled channels which are generated between the nodes $"1"\&"5"
(\rho_{15})$ and $"1"\&"6" (\rho_{16})$. In general the degree of
entanglement is much smaller than that depicted Sec.3. The
location of the node in the network has a noticeable effect on the
 entanglement value, where for nearer nodes the entanglement is
much larger than that displayed for the long distance locations.
Also, the smooth behavior of entanglement is clear for the nearer
nodes, while the phenomena of the sudden death and birth is
depicted for entanglement which is generated between long distance
nodes.

\section{Teleportation}
In this section, we investigate the possibility of using the
generated entangled states between different nodes to communicate
among themselves. For this aim we assume that the node $"1"$ is
given unknown state defined by
$\ket{\psi}_u=\alpha\ket{0}+\beta\ket{1}$. The density operator of
the state is given by,
\begin{equation}\label{Unk}
\rho_u=\frac{1}{2}(1+s_{u_x}\sigma_x+s_{u_y}\sigma_y+s_{u_z}),
\end{equation}
where $s_{u_x}=\alpha\beta^*+\alpha^*\beta,
s_{u_y}=i(\alpha^*\beta-\alpha\beta^*)$ and
$s_{u_z}=|\alpha|^2-|\beta|^2$. Then the total state of the
network  is $\rho_{s}=\rho_u\otimes\rho_{ij}$ where $ij=12,13,14$.
Now to teleport this state to the nodes $2$, $3$ or $4$, theses
nodes perform the following steps \cite{Ben}:
\begin{figure}[b!]
  \begin{center}
 \includegraphics[width=25pc,height=15pc]{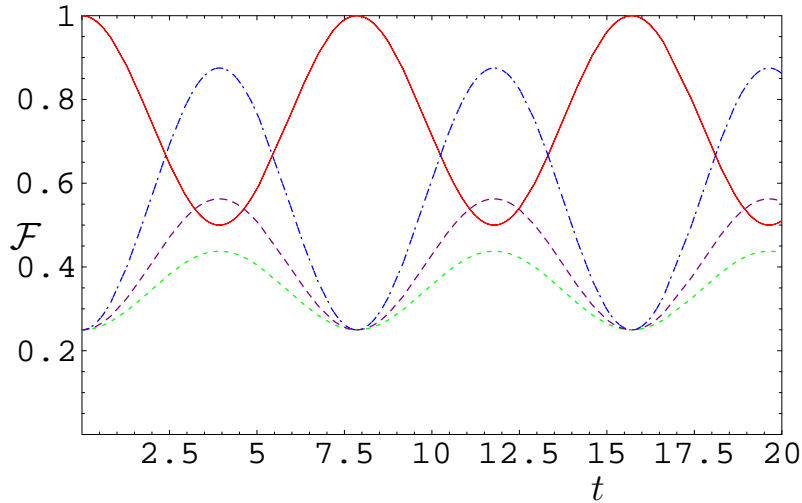}~\quad
  \put(-80,-10){\large $t$}
 \put(-300,85){\large $\mathcal{F}$}
    \caption{The fidelity of the teleported state from between different
    channels in the netork.The solid curve represents fidelity of
    telepoted state from the node 1 to the node $2$ via the channel
    $\rho_{12}$. The dash-dot curve represents  $\mathcal{F}$ via
    the channel $\rho_{14}$. Finally the dot curve represents the
 $\mathcal{F}$ via the channel $\rho_{23}$, where DM interaction is
 considered in the $x$-direction.}
     \end{center}
\end{figure}

\begin{enumerate}
\item The first node performs the CNOT operation between its own
qubit and the qubit  which is located at  nodes $2, 3$ or $4$.
\item The first node applies the Hadamard on its own particle.
\item Then the first node's qubit and one of the other qubits are
measured  randomly in one of the Bell states $\rho_{\phi^\pm}$ or
$\rho_{\psi^\pm}$. Then the teleported state takes the form,
\begin{equation}\label{rt}
\rho_{t}=\frac{1}{2}(1+s_{t_x}\sigma_x+s_{t_y}\sigma_y+s_{t_z}\sigma_z),
\end{equation}
where $s_{t_x}, s_{t_y},s_{t_z}$  are the Bloch vector for the
final teleported state. To quantify the closeness of the input
state (\ref{Unk}) with the final state (\ref{rt}), we evaluate the
fidelity $\mathcal{F}$ which is defined as,
\begin{equation}
\mathcal{F}=\frac{1}{4}(1+s_{u_x}s_{t_x}+s_{u_yx}s_{t_y}+s_{u_z}s_{t_z}).
\end{equation}
\end{enumerate}

In Fig.(8), we plot the Fidelity of the teleported state from the
node $1$ to the node $2$. It is clear that at $t=0$, the fidelity
is maximum $(\mathcal{F}=1$). As soon as the interaction is going
on, the fidelity decreases to reach its minimum value ($=0.45)$ at
$t\simeq 2.6$.  Then the $\mathcal {F}$ re-increases to  become
maximum at $\simeq 5$. This behavior is repeated periodically
depending on the value of the interaction strength, where as $D_x$
increases the number of repletion increases. The fidelity of
teleportating the unknown state to the node $3$ is defined by the
dot curve in Fig.(8). At $t=0$, the fidelity is very small  due to
the classical correlation and it is called classical
teleportation. For $t>0$, there is an entangled state  generated
between the  nodes $2$ and $3$ and hence the fidelity increases to
reach its maximum value at $t\simeq 2.6$. However for larger $t$,
the fidelity decreases( due to the loss of entanglement between
the two nodes) and reaches its minimum value at $t=5$.  The same
behavior is seen for the fidelity of the teleported state between
the nodes $1$ and $4$ (dash-dot curve). It shows that the behavior
is the same as that depicted between the nodes (2$\&$ 3), but the
maximum values are smaller.

\section{conclusion}
In this contribution, an entangled network is constructed by using
maximum entangled pairs. These pairs, which represent the  nodes
of the network,  interact together via  DM interaction, where we
consider that DM  interaction is switch on the  $z-$ or $x-$
directions. Due to this interaction, there are entangled channels
 generated between these nodes. The amount of  entanglement which
is contained in these channels is quantified by using Wootters
concurrence. It is shown that  the phenomena of the sudden death
and re-birth  of entanglement  appear for these channels which are
generated via indirect interaction. However, the concurrence
increases and deceases smoothly for those channels which are
generated via direct interaction. The strength of DM interaction
plays an important role on the period of death - rebirth
entanglement. Also, for nearer nodes the entangled channels are
generated much faster than those located in a long distance.

The amount of entanglement between different channels depends on
the direction of DM interaction. We show that when DM is switched
on the $x$ axis, there are maximum entangled channels are
generated between some nodes and the entanglement of initial
entangled nodes doesn't vanish. However if DM interaction  is
switched in the  $z-$ axis, there is no maximum entangled states
between any two nodes  generated via indirect interaction while
the initial entangled channels loose their entanglement very fast.
The minimum amount of entanglement  contained in the four-nodes
network is quantified. The upper and lower bounds of the
entanglement of the generated network depends on the direction of
DM interaction and the repatation of the behavior depends on the
strength of DM.

Finally, the quantum entangled channels between the different
nodes are used to perform quantum teleportation. We clarify this
idea by considering  the entangled channels  generated between
different nodes when DM  interaction is switched on the $x-$axis.
It is clear that the fidelity of the teleported state depends on
the location between the nodes. For those initially entangled, the
fidelity decreases smoothly but doesn't vanishe.

\end{document}